\documentclass[prd,onecolumn,nofootinbib,showpacs]{revtex4}
\usepackage{epsfig,graphicx,bm}
\usepackage[latin2]{inputenc}
\usepackage{color}
\usepackage{t1enc}
\usepackage{amssymb}
\usepackage{amsmath}
\usepackage{graphicx}
\usepackage{setspace}

\newcommand{\tr}{\textnormal{Tr\,}}
\newcommand{\bea}{\begin{eqnarray}}
\newcommand{\eea}{\end{eqnarray}}
\newcommand{\be}{\begin{equation}}
\newcommand{\ee}{\end{equation}}
\newcommand{\slk}{\mbox{\,\slash \hspace{-0.5em}$k$}}

\newcommand{\slparc}{\mbox{$\,\slash$ \hspace{-0.9em}$\partial$}}
\renewcommand{\k}{{\bf k}}
\newcommand{\q}{{\bf q}}
\newcommand{\p}{{\bf p}}
\def\lag{\langle}
\def\rag{\rangle}
\newcommand{\no}{{\nonumber}}

\newcommand{\Pval}{\mathop{\mathcal P}}

\newcommand{\rt}[1]{{}}

\begin{document}
{\allowdisplaybreaks

\title{
Influence of the isospin and hypercharge chemical potentials
on the location of the CEP in the $\mu_B-T$ phase diagram  of the
$SU(3)_L \times SU(3)_R$ chiral quark model
}

\author{P. Kov\'acs}
\email{kpeti@cleopatra.elte.hu}
\affiliation{Research Group for Statistical and Biological Physics 
of the Hungarian Academy of Sciences,
H-1117 Budapest, Hungary}

\author{Zs. Sz{\'e}p}
\email{szepzs@achilles.elte.hu}
\affiliation{Research Institute for Solid State Physics and Optics
of the Hungarian Academy of Sciences, H-1525 Budapest, Hungary}

\begin{abstract}
We investigate the influence of the asymmetric quark matter
($\rho_u\ne\rho_d\ne\rho_s$) on the mass of the
quasiparticles and the phase diagram of the chiral quark model
parametrized at one-loop level of the renormalized theory, using the
optimized perturbation theory for the resummation of the perturbative
series. The effect of various chemical potentials introduced in the
grand canonical ensemble is investigated with the method of
relativistic many-body theory. The temperature dependence of the
topological susceptibility is estimated with the help of the
Witten--Veneziano mass formula.
\end{abstract}


\maketitle

\section{Introduction}

The study of a system of particles at finite density and temperature
is phenomenologically interesting because in heavy ion collision
experiments the initial state is such that the chemical potentials
$\mu_B,\mu_I, \mu_Y$ (conjugate to the baryon charge, third component
of the isospin and hypercharge, respectively)  are non vanishing,
although the last two are much smaller than the first one. Assuming
thermal equilibrium, thermal models show that the strangeness chemical
potential in central Si$+$Au collisions at the Brookhaven AGS
experiment was 20-25\% of the baryonic chemical potential for which
the best fit gives $\mu_B=540$~MeV \cite{braun-munzinger95}. For
central Pb$+$Pb collisions at CERN SPS experiments the value of the
strangeness chemical potential was $\sim25-30\%$ and that of the
isospin chemical potential $\sim2-5\%$ of the value of $\mu_B$
estimated to be around $233-266$~MeV
\cite{braun-munzinger99,becattini98}. The Compressed Baryonic Matter
(CBM) experiment at FAIR in Darmstadt will explore regions of the QCD
phase diagram with moderate temperature up to such high values of the
baryonic density which are comparable with those in the core of
neutron stars~\cite{CBM06}.

In many-body theory chemical potential is introduced to any
conserved charge. In heavy ion collision experiments the baryon
number, isospin and hypercharge can be considered conserved due to the
short time elapsed between the formation of the fireball and its
freeze-out, during which only the strong interactions play 
important role, the electroweak interactions being negligible. It
is expected that in the very early stage of the fireball's evolution
strangeness is abundantly produced in the deconfined phase through
gluon-gluon fusion \cite{rafelski82}, while in the hadronic phase in
the vicinity of the transition multi-mesonic 
reactions will play an important role in the fast redistribution of
strange quarks~\cite{greiner01}. 

The influence of the isospin chemical potential on the chiral phase
transition is currently actively investigated, because this effect can
in principle be tested experimentally. As noticed in
\cite{toublan05a}, using different isotopes of an element in heavy ion
collision experiments will vary $\mu_I$ keeping $\mu_B$
constant. Moreover, the system with real $\mu_I$ represents no extra
difficulty in lattice field theory compared to the introduction of
$\mu_B$. For two flavors the simulations at $\mu_B=0$ and $\mu_I\ne0$
is not even affected by the sign problem \cite{alford99}.  For
$\mu_I\ne 0$ a generic result coming from effective models of the
strongly interacting matter without the $U(1)_A$ anomaly appeared to
be the splitting in the $\mu_B-T$ plane of the first order transition
line into two transition lines. This effect was observed in random
matrix model \cite{klein03}, NJL model \cite{toublan03}, strong
coupling limit of the staggered lattice QCD \cite{nishida}, all with
two flavors and in the three flavors ladder QCD \cite{barducci03}.
This would imply the existence of not only the two phases having
$\lag\bar u u\rag\ne 0$, $\lag\bar d d\rag\ne 0$ and $\lag\bar u
u\rag=\lag\bar d d\rag =0$, respectively, but also of a phase with
$\lag\bar u u\rag=0$ and $\lag\bar d d\rag \ne 0$. It was shown in
\cite{frank03,he05a} that the structure with two transition lines and
critical end points ceases to exist for a sufficiently strong $U(1)_A$
breaking, above which the two strongly coupled condensates vanish
simultaneously.  In a hadron resonance gas model it was found that at
fixed baryon chemical potential the pseudocritical temperature of the
transition between the hadronic and the quark-gluon plasma phases is
lowered as either the isospin or the strangeness chemical potential is
increased \cite{toublan05b}.

Because of their phenomenological implications, it is natural to study
to what extent these results are present in another low energy
effective model, the chiral quark model, widely used for studying the
chiral behavior of strongly interacting matter. In the past few years
we have investigated the thermodynamics of this model for two and three
quark flavors at $\mu_B=0$ and $\mu_B\ne 0$, while leaving
$\mu_I=\mu_Y=0$ \cite{jpszsz04,herpay05,herpay06,kovacs07}.  As a
continuation of these previous studies, in this paper we consider the
influence of the chemical potentials on the chiral phase transition up
to such high values of the isospin chemical potential above which the
condensation of pseudoscalar mesons occurs. The pion and kaon
condensation phase, which is beyond the scope of our present
investigation, was studied both with lattice methods and using
effective theories~\cite{son00,kogut03,sinclair04,he05b,warringa05,sinclair06,andersen07}.

The paper is organized as follows. In section~\ref{sec:model} we
present the model, its one-loop parametrization and the introduction
of the chemical potentials. The variation of the location of the
critical end point in presence of $\mu_I$ and $\mu_Y$ is studied in
section~\ref{sec:thermo}. There we investigate also the temperature
and density dependence of the one-loop pole masses of the pseudoscalar
mesons. We conclude in section~\ref{sec:conclusion}.

\section{The $SU(3)_L\times SU(3)_R$ symmetric chiral quark model 
\label{sec:model}}

The Lagrangian of the model containing explicit symmetry breaking terms is
\bea
\nonumber
L&=&\frac{1}{2}\tr(\partial_\mu M^{\dag} \partial^\mu M+
m_0^2 M^{\dag} M)-
f_1 \left( \tr(M^{\dag} M)\right)^2-f_2  \tr(M^{\dag} M)^2-
g\left(\det(M)+\det(M^{\dag})\right)+
\epsilon_0\sigma_0+\epsilon_3\sigma+\epsilon_8 \sigma_8
\\&&
+\bar\psi\left(i\slparc-g_F M_5\right)\psi.
\label{Lagrangian}
\eea

The constituent quarks are contained in the field $\psi$:
$\bar\psi=(\bar u,\bar d,\bar s)$.  The two $3\times3$ complex
matrices are defined in terms of the scalar $\sigma_i$ and
pseudoscalar $\pi_i$ fields as
$M~=~\frac{1}{\sqrt{2}}\sum_{i=0}^{8}(\sigma_i+i\pi_i)\lambda_i$ and
$M_5~=~\frac{1}{2}\sum_{i=0}^{8}(\sigma_i+i\gamma_5\pi_i)\lambda_i$,
with $\lambda_i\,:\,\,\,i=1\ldots 8$ the Gell--Mann matrices and
$\lambda_0:=\sqrt{\frac{2}{3}} {\bf 1}.$ The fields with well defined
quantum numbers are obtained with a block-diagonal transformation
$f_\alpha=T_{\alpha i}f_i$, $f\in\{\sigma,\pi\}$, where
$T=\textrm{diag}(1,\tau,1,\tau,\tau,1)$ and
$\tau=\frac{1}{\sqrt{2}}\begin{pmatrix}1 & -i\\ 1 &i\end{pmatrix}$.
As $\alpha$ goes from $0$ to $8$, the components of the scalar and
pseudoscalar fields go trough $\sigma_0,a_0^+,a_0^-,\sigma_3,$
$\kappa^+,\kappa^-,\kappa^0,\bar\kappa^0,\sigma_8$ and
$\pi_0,\pi^+,\pi^-,\pi_3,$ $K^+,K^-,K^0,\bar K^0,\pi_8$,
respectively. The physical fields $\pi^0$ (neutral pion), $\eta$ and
$\eta'$ mesons in the pseudoscalar sector and $a_0^0$ (neutral $a_0$),
$\sigma$ and $f_0$ in the scalar sector are obtained as linear
combinations of the corresponding fields in the two mixing 0,3,8 sectors.

In this paper we investigate the pattern of symmetry breaking realized
in nature, with the $SU(3)_A\times U(1)_A\times SU(3)_V$ symmetry
completely broken, that is the isospin $SU(2)_V$ is also broken. In
addition to the spontaneous symmetry breaking, explicit breaking is
also considered with the introduction of external fields for all the
diagonal generators of the scalar sector. This results in having three
non-vanishing condensates in the broken symmetry phase:
$v_{\delta}=\lag \sigma_{\delta} \rag$, for $\delta=0,3,8$. The
condensates determines the tree-level scalar and pseudoscalar masses:
\be
\begin{split}
&m_{S,\alpha\beta}^2 = m^2 \delta_{\alpha\beta} 
- 6 \tilde G_{\alpha\beta\gamma} v_\gamma
+ 4 \tilde F_{\alpha\beta\gamma\delta} v_\gamma v_\delta,
\\
&m_{P,\alpha\beta}^2 = m^2 \delta_{\alpha\beta} 
+ 6 \tilde G_{\alpha\beta\gamma} v_\gamma
+ 4 \tilde H_{\alpha\beta,\gamma\delta} v_\gamma v_\delta. 
\end{split}
\ee
The tensors appearing above arise after the evaluation of the trace
in (\ref{Lagrangian}) and the transformation of the fields to 
the basis with good quantum numbers.
The connection between these coupling tensors and the original ones 
appearing in (\ref{Lagrangian}) which
can be found in \cite{Haymaker73,Lenaghan00} is given by:
\be
\tilde{G}_{\alpha \beta\gamma}=
\sum_{i,j,k=0}^8G_{ijk}T^{-1}_{i\alpha}T^{-1}_{j\beta}T^{-1}_{k\gamma},\quad
\tilde{H}_{\alpha \beta,\gamma \delta}=
\sum_{i,j,k,l=0}^8H_{ij,kl}{T}^{-1}_{i\alpha}T^{-1}_{j\beta}
T^{-1}_{k\gamma}T^{-1}_{l\gamma},\quad
\tilde{F}_{\alpha \beta \gamma \delta}=
\sum_{i,j,k,l=0}^8F_{ijkl}T^{-1}_{i\alpha}T^{-1}_{j\beta}
T^{-1}_{k\gamma}T^{-1}_{l\gamma}.
\label{Eq:tilde_coeffs}
\ee
The transformations preserve the symmetry structure of the tensors,
that is $\tilde{G}_{\alpha \beta\gamma}$ and 
$\tilde{F}_{\alpha \beta \gamma \delta}$ are completely symmetric and
$\tilde{H}_{\alpha \beta,\gamma \delta}$ is symmetric upon the interchange
of two indices which are on the same side of the comma.

The tree-level mass square matrices are not diagonal in the $0,3,8$ subspace,
but since they are real and symmetric diagonalization is achieved with
an orthogonal transformation. The tree-level orthogonal matrices in
the scalar and pseudoscalar sectors are denoted with $O_S$ and $O_P$,
respectively. Denoting the eigenvalues of the pseudoscalar and the
scalar $3\times3$ mass matrices in the $0,3,8$ sector with
$\lambda_{P,\{\text{min,mid,max}\}}$ and
$\lambda_{S,\{\text{min,mid,max}\}}$ the tree-level masses of the
mesons are as follows 
\vspace*{-0.5cm}
\begin{spacing}{1.2}
\bea
\begin{array}{ll}
m_{\pi^+}^2=m_{\pi^-}^2=m_{P,12}^2,\quad&
m_{a_0^+}^2=m_{a_0^-}^2=m_{S,12}^2,\\
m_{\pi^0}^2=\lambda_{P,\text{min}},\quad &
m_{a_0^0}^2=\lambda_{S,\text{mid}}, \\
m_{K^+}^2=m_{K^-}^2=m_{P,45}^2,\quad &
m_{\kappa^+}^2=m_{\kappa^-}^2=m_{S,45}^2, \\
m_{K^0}^2=m_{\bar K^0}^2=m_{P,67}^2,\quad &
m_{\kappa^0}^2=m_{\bar\kappa^0}^2=m_{S,67}^2, \\
m_{\eta}^2=\lambda_{P,\text{mid}}, \quad &
m_{\sigma}^2=\lambda_{S,\text{min}}, \\
m_{\eta^{\prime}}^2=\lambda_{P,\text{max}}, \quad &
m_{f_0}^2=\lambda_{S,\text{max}}.
\end{array}
\label{Eq:m_tree}
\eea
\end{spacing}
Note, that some of the tree-level masses of scalars and pseudoscalars
coincide. As we will see, the introduction of isospin and hypercharge
chemical potentials distinguish between the particles and as a result
all the one-loop pole masses will be different for $\mu_I,\mu_Y\ne 0$.

The tree-level fermion masses are: 
\be
M_u=\frac{g_F}{\sqrt{12}}(\sqrt{2}v_0+\sqrt{3}v_3+v_8),\qquad
M_d=\frac{g_F}{\sqrt{12}}(\sqrt{2}v_0-\sqrt{3}v_3+v_8),\qquad
M_s=\frac{g_F}{\sqrt{12}}(\sqrt{2}v_0-2v_8).
\label{Eq:fermion_masses}
\ee

The evolution of the condensates with the temperature or/and the
chemical potentials is determined by the three equations of state
\bea
\nonumber
\label{Eq:eos_0}
0&=&\lag \frac{\partial L}{\partial\sigma_0} \rag=
m^2v_0-\frac{c}{2\sqrt{6}}(2v_0^2-v_3^2-v_8^2)+
\frac{1}{3}(3g_1+g_2)v_0^3+(g_1+g_2)(v_3^2+v_8^2)v_0+
\frac{g_2}{3\sqrt{2}}(3v_3^2-v_8^2)v_8-\varepsilon_0\nonumber\\
&-&\sum_{\substack{f\in\{\sigma,\pi\} \\ \alpha=1,2,4\dots7}}\!\!
t^0_{f,\alpha} \lag f_{\alpha}^\dag f_{\alpha} \rag
-3\sum_{\gamma\in\{0,3,8\}}\!\!
\left[\left(O_S^T S_0 O_S\right)_{\gamma\gamma}\lag \sigma_{\gamma} \sigma_{\gamma} \rag
+\left(O_P^T P_0 O_P\right)_{\gamma\gamma}\lag \pi_{\gamma} \pi_{\gamma} \rag\right]
+\frac{g_F}{\sqrt{6}}
N_c(\lag\bar u u\rag+\lag\bar d d\rag+\lag\bar s s\rag),
\\
\label{Eq:eos_3}
0&=&\lag \frac{\partial L}{\partial\sigma_3} \rag=
\left(m^2-\frac{c}{\sqrt{3}}v_8+\frac{c}{\sqrt{6}}v_0+
(g_1+\frac{g_2}{2})(v_3^2+v_8^2)+(g_1+g_2)v_0^2+\sqrt{2}g_2v_0v_8\right)v_3
-\epsilon_3
\nonumber\\
&-&\sum_{\substack{f\in\{\sigma,\pi\} \\ \alpha=1,2,4\dots7}} t^3_{f,\alpha}
\lag f_{\alpha}^\dag f_{\alpha} \rag
-3\sum_{\gamma\in\{0,3,8\}}
\left[\left(O_S^T S_3 O_S\right)_{\gamma\gamma} \lag \sigma_{\gamma} \sigma_{\gamma} \rag
+\left(O_P^T P_3 O_P\right)_{\gamma\gamma}\lag \pi_{\gamma} \pi_{\gamma} \rag\right]
+\frac{g_F}{2}N_c(\lag\bar u u\rag-\lag\bar d d\rag),
\\
\nonumber
\label{Eq:eos_8}
0&=&\lag \frac{\partial L}{\partial\sigma_8} \rag=
m^2v_8+\frac{c}{\sqrt{6}}v_0v_8+\frac{c}{2\sqrt{3}}(v_8^2-v_3^2)+
(g_1+\frac{g_2}{2})(v_8^2+v_3^2)v_8+\frac{g_2}{\sqrt{2}}(v_3^2-v_8^2)v_0+
(g_1+g_2)v_0^2v_8-\varepsilon_8
\nonumber\\
&-&\sum_{\substack{f\in\{\sigma,\pi\} \\ \alpha=1,2,4\dots7}} t^8_{f,\alpha}
\lag f_{\alpha}^\dag f_{\alpha} \rag
-3\sum_{\gamma\in\{0,3,8\}}
\left[\left(O_S^T S_8 O_S\right)_{\gamma\gamma} \lag \sigma_{\gamma} \sigma_{\gamma} \rag
+\left(O_P^T P_8 O_P\right)_{\gamma\gamma}\lag \pi_{\gamma} \pi_{\gamma} \rag \right]\nonumber\\
&+&\frac{g_F}{2\sqrt{3}}
N_c(\lag\bar u u\rag+\lag\bar d d\rag-2\lag\bar s s\rag),
\eea
where in the mixing sector $\sigma_\gamma$ stands for
$\sigma,a_0^0,f_0$ and similarly $\pi_\gamma$ denotes
$\pi^0,\eta,\eta^{\prime}$ as $\gamma=0,3,8$,
respectively. $f_{\alpha}^\dag$ denotes the antiparticle of
$f_{\alpha}$, that is
{\it e.g.} for $f=\sigma$ and $\alpha=1$ one has $\sigma_1=a_0^+$ and 
$\sigma_1^\dag=a_0^-$. In this notation
$\lag f_{\alpha}^+ f_{\alpha}
\rag=T_{B}^{\beta}(m_{f_{\alpha}})$, $\lag \bar q q
\rag=-4m_{q}T_{F}^{\beta}(m_{q})$, where
$T_{B}^{\beta}(m_{f_{\alpha}})$ and $T_{F}^{\beta}(m_{q})$ stands for the
bosonic, and the fermionic tadpole integrals, respectively. These
integrals are given in Appendix B of \cite{kovacs07}.
The coefficients
$t^{\gamma}_{f,\alpha}$ are listed in Table~\ref{Tab:t_coeff}.
In the mixing sector, that is for $\gamma=0,3,8$, the $3\times 3$ matrices
read:
\be
\begin{split}
&S_{\gamma}=\tilde{G}_0-\frac{4}{3}v_0\tilde{F}_{\gamma 0}-
\frac{4}{3}v_3\tilde{F}_{\gamma 3}-\frac{4}{3}v_8\tilde{F}_{\gamma 8},\\
&P_{\gamma}=\tilde{G}_0+\frac{4}{3}v_0\tilde{H}_{\gamma 0}+
\frac{4}{3}v_3\tilde{H}_{\gamma 3}+\frac{4}{3}v_8\tilde{H}_{\gamma 8},
\end{split}
\ee
with the definition: 
$(\tilde{G}_{\gamma})_{\alpha\beta}\equiv \tilde{G}_{\gamma\alpha\beta}$, 
$(\tilde{F}_{\gamma \delta})_{\alpha\beta}\equiv 
\tilde{F}_{\alpha\beta\gamma\delta}$,
and 
$(\tilde{H}_{\gamma \delta})_{\alpha\beta}\equiv 
\tilde{H}_{\alpha\beta,\gamma\delta}$. All the indices run through  0, 3, or
8.

\begin{table}
\[
\begin{array}{l|c||l|c}  
  \alpha & t^{\gamma}_{\sigma,\alpha} & \alpha & t^{\gamma}_{\pi,\alpha} \\ 
  \hline
   a_0^- &
   3\tilde{G}_{\gamma21}-4\tilde{H}_{\gamma21\delta}v_{\delta}
   & \pi^- & 
   3\tilde{G}_{\gamma21}-4\tilde{F}_{\gamma21\delta}v_{\delta} \\
   a_0^+ & 
   3\tilde{G}_{\gamma12}-4\tilde{H}_{\gamma12\delta}v_{\delta}
   & \pi^+ &
   3\tilde{G}_{\gamma12}-4\tilde{F}_{\gamma12\delta}v_{\delta} \\
   \kappa^- &
   3\tilde{G}_{\gamma54}-4\tilde{H}_{\gamma54\delta}v_{\delta}
   & K^- & 
   3\tilde{G}_{\gamma54}-4\tilde{F}_{\gamma54\delta}v_{\delta} \\
   \kappa^+ &
   3\tilde{G}_{\gamma45}-4\tilde{H}_{\gamma45\delta}v_{\delta}
   & K^+ & 
   3\tilde{G}_{\gamma45}-4\tilde{F}_{\gamma45\delta}v_{\delta} \\
   \bar{\kappa}^0 &
   3\tilde{G}_{\gamma76}-4\tilde{H}_{\gamma76\delta}v_{\delta}
   & \bar{K}^0 & 
   3\tilde{G}_{\gamma76}-4\tilde{F}_{\gamma76\delta}v_{\delta} \\
   \kappa^0 & 
   3\tilde{G}_{\gamma67}-4\tilde{H}_{\gamma67\delta}v_{\delta}
   & K^0 & 
   3\tilde{G}_{\gamma67}-4\tilde{F}_{\gamma67\delta}v_{\delta}                  
\end{array} 
.
\]
\caption{The $t^{\gamma}_{f,\alpha}$ coefficients appearing in the equations of
state (\ref{Eq:eos_0}), (\ref{Eq:eos_3}), (\ref{Eq:eos_8}). The
summation index $\delta$ goes over $0,3,8$.}
\label{Tab:t_coeff}
\end{table}

\subsection{One-loop parametrization of the model at 
zero temperature and density }

One has some freedom in choosing the set of equations which determines
the 13 parameters of the model, namely
$m_0^2,f_1,f_2,g,g_F,$\,$v_0,v_3,v_8,$\,$\epsilon_0,\epsilon_3,\epsilon_8$
and $l_f,l_b$. These latter two parameters are the fermionic and
bosonic renormalization scales. For the parametrization we follow the
method described in \cite{kovacs07} where the renormalization of the
model was also discussed. The only difference in the present case is
the appearance of $v_3$ and $\epsilon_3$. Since at zero temperature
and densities the effect of isospin breaking is small we use the same
values for $l_f$ and $l_b$ as in \cite{kovacs07} where these were
determined by minimizing the deviation of the predicted mass spectrum
from the physical one. The external fields are determined from the
equations of state (\ref{Eq:eos_0}), (\ref{Eq:eos_3}),
(\ref{Eq:eos_8}) once the remaining 8 parameters are known.

In order to avoid the appearance of negative propagator mass squares
in the one-loop finite temperature calculations in the broken symmetry
phase we use the Optimized Perturbation Theory (OPT) of
Ref.~\cite{hatsuda98}. This amounts to replace the mass parameter
$-m_0^2$ in the Lagrangian with an effective, eventually 
temperature-dependent, mass parameter $m^2$:
\be
L_{mass}=\frac{1}{2}m^2 \tr M^\dag M-\frac{1}{2}(m_0^2+m^2) \tr M^\dag M
\equiv\frac{1}{2}m^2 \tr M^\dag M-\frac{1}{2} \Delta m^2\tr M^\dag M.
\label{resum}
\ee
The counterterm $\Delta m^2$ is taken into account first at one-loop
level, while $m^2$ will replace $m_0^2$ in all the tree-level masses and is
determined using the criterion of fastest apparent convergence
(FAC). We have chosen to implement this criterion by requiring that
for $\pi^+$ the one-loop mass calculated at vanishing external
momentum stays equal to the tree-level mass ($M_{\pi^+}=m_{\pi^+}$).
We have checked that imposing this equation for the neutral pion
rather than the charged one results in no significant changes in the
parameters. We note here that we were forced to use the definition
$M_{\pi^+}^2=-iG^{-1}(p=0)$ instead of defining the one-loop mass as
the pole of the propagator because in this latter case the solution to
the gap equation, to be presented below, ceases to exist above a
certain temperature, in accordance with previous investigations using
the OPT \cite{hatsuda98,herpay06}. 

\begin{figure}[t]
\centerline{\includegraphics[keepaspectratio,width=1.0\textwidth,angle=0]{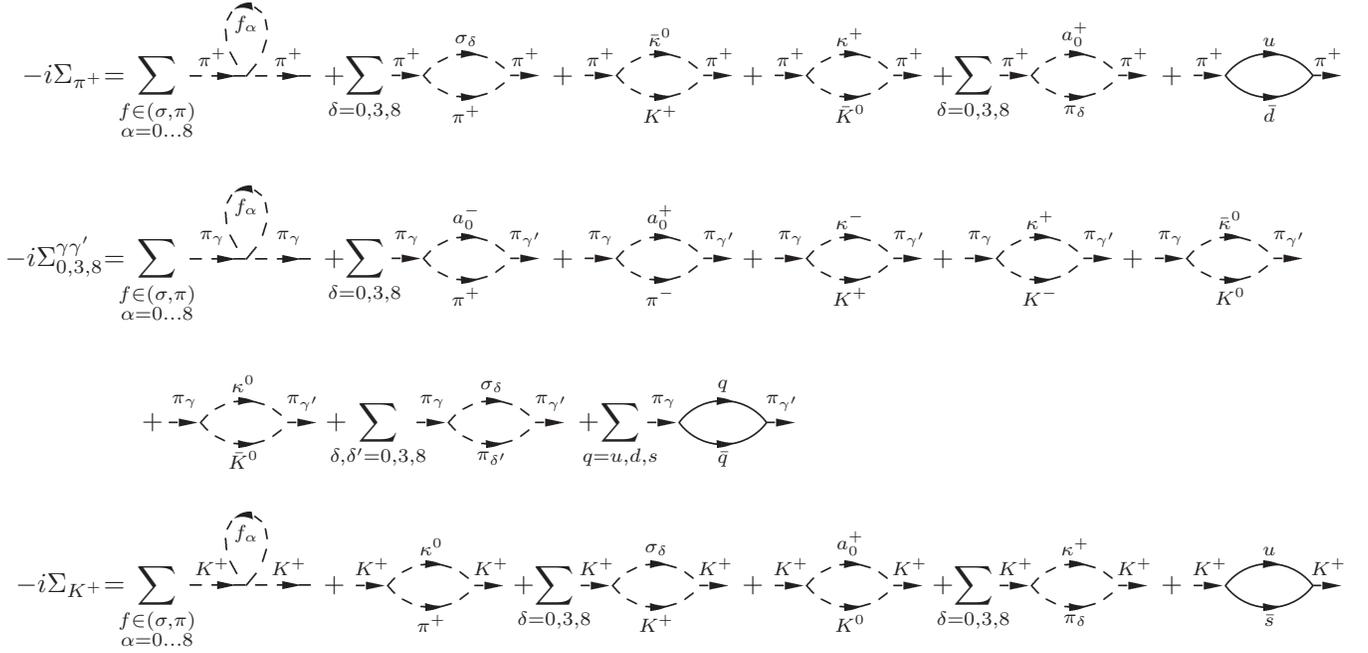}
}
\caption{
Diagrammatic representation of the one-loop pseudoscalar self-energies used
for the parametrization. The label associated to the line denotes the
propagating particle.}
\label{Fig:self-energies}
\end{figure}

As described in details in \cite{kovacs07}, with the application of
FAC one can eliminate the effective mass parameter $m^2$ in favor of
the tree-level pion mass $m_{\pi^+}^2$ in all the other tree-level
masses of the propagators used to calculate the one-loop
self-energies:
\be
m^2 =
m_{\pi^+}^2+\frac{c}{\sqrt{6}}v_0-\frac{c}{\sqrt{3}}v_8-
\frac{\sqrt{2}}{3}g_2v_0v_8-\left(g_1+\frac{g_2}{3}\right)v_0^2-
\left(g_1+\frac{3g_2}{2}\right)v_3^2-\left(g1+\frac{g_2}{6}\right)v_8^2.
\ee
In this way one obtains the following gap equation
\be
m_{\pi^+}^2=-m_0^2-\frac{c}{\sqrt{6}}v_0+\frac{c}{\sqrt{3}}v_8+
\frac{\sqrt{2}}{3}g_2v_0v_8+\left(g_1+\frac{g_2}{3}\right)v_0^2
+\left(g_1+\frac{3g_2}{2}\right)v_3^2+\left(g_1+\frac{g_2}{6}\right)v_8^2+
\textnormal{Re}\Sigma_{\pi^+}(p^2=0,m_i^2(m_{\pi^+}^2)),
\label{Eq:pi+_gap_eq}
\ee
where $\Sigma_{\pi^+}$ denotes the self-energy of $\pi^+$ shown
diagrammatically in Fig.~{\ref{Fig:self-energies}}. 
Equation (\ref{Eq:pi+_gap_eq}) is the first from a set of four coupled
non-linear equations which determines $m_0^2,f_1,f_2,g$, if one knows
$g_F,v_0,v_3,v_8$. Two further equations of the set are given by the
one-loop equation for the $\eta$ and $K^+$ pole masses
\bea
M_{\eta}^2&=&-m_0^2+\left[\tilde{O}_P^T\Big({\mathcal M}^2_{\textrm{tree}}+
\textnormal{Re}\Sigma_{0,3,8}(p^2=M_{\eta}^2)\Big)
\tilde{O}_P\right]_{22},\\
M_{K^+}^2&=&-m_0^2-\frac{c}{\sqrt{6}}v_0+\frac{c}{2}v_3-
\frac{c}{2\sqrt{3}}v_8+\frac{g_2}{\sqrt{6}}v_0v_3-
\frac{g_2}{3\sqrt{2}}v_0v_8+\frac{2g_2}{\sqrt{3}}v_3v_8\nonumber\\
&&+\left(g_1+\frac{g_2}{3}\right)v_0^2+\left(g_1+\frac{g_2}{2}\right)v_3^2+
\left(g_1+\frac{7g_2}{6}\right)v_8^2+
\textnormal{Re}\Sigma_{K^+}(p^2=M_{K^+}^2),
\eea
where ${\mathcal M}^2_{\textrm{tree}}$ is the tree-level mass squared 
matrix of the mixing sector without the mass parameter $m^2$, the orthogonal
matrix $\tilde{O}_P$ diagonalizes the expression in the round bracket,
and $\Sigma_{0,3,8}$ is the self-energy matrix of the pseudoscalar
mixing sector. This matrix is determined numerically.
 The last equation in the set is the FAC criterion for
the kaon, which requires 
$\textnormal{Re}\Sigma_{K^+}(p^2=M_{K^+}^2)-\Delta m^2=0$.

The parameters $g_F,v_0,v_3,v_8$ are determined as follows. 
A linear combination of
$v_0$ and $v_8$ is determined by the tree-level PCAC relation for the
pion decay constant (see Appendix of \cite{Lenaghan00})
\be
f_\pi:=d_{11a}v_a=\sqrt{\frac{2}{3}}v_0 +\frac{1}{\sqrt{3}}v_8.
\label{Eq:PCAC}
\ee 
One can see from (\ref{Eq:fermion_masses}) that the same linear
combination enters the expression of the average mass of the two
light constituent quarks, so that the Yukawa coupling is given by
$g_F=(M_u+M_d)/f_\pi$.  Another linear combination of $v_0$ and $v_8$
appears in the expression of $M_s$ in (\ref{Eq:fermion_masses}) which
together with the PCAC relation (\ref{Eq:PCAC}) determines $v_0$ and
$v_8$:
\be
v_0 = \sqrt{\frac{2}{3}}f_{\pi}\left(1+\frac{M_s}{M_u+M_d}\right),\qquad 
v_8 = \frac{1}{\sqrt{3}}f_{\pi}\left(1-\frac{2M_s}{M_u+M_d}\right).
\ee  
The remaining parameter, $v_3$ is obtained by requiring that
the difference between the tree-level masses of $\pi^+$ and $\pi^0$
equals the physical value ($\Delta m_{\pi}$):
\be
-\left[O_P^T{\mathcal M}^2_{\textrm{tree}}O_P\right]_{11}=
\left(\Delta m_{\pi}\right)^2.
\ee

This equation has two roots for $v_3$, a negative and a positive one.
The positive root would give $m_{K^0} < m_{K^+}$ for the kaon
masses. Since the opposite relation holds in nature, we choose the
negative solution which is the physically valid one.

We use the following values for the physical quantities: 
$m_{\pi^+}=139.57$~MeV, $\Delta m_{\pi}=4.594$~MeV, $M_{K^+}=493.677$~MeV,
$M_\eta=547.8$~MeV, $f_\pi=93$~MeV, $(M_u+M_d)/2=313$~MeV, $M_s=530$~MeV
and in addition $l_b=520$~MeV and $l_f=1210$~MeV for the two
renormalization scales.  
   
\subsection{Introduction of the chemical potentials\label{chem_intro}}

The introduction of the chemical potential for a system with a set of
conserved charge operators is reviewed below.  For vanishing external
fields the Lagrangian (\ref{Lagrangian}) is invariant under the
following global vector transformations
\be
\begin{split}
&
M\to e^{-i\alpha_G G} M e^{i\alpha_G G}=M-i\alpha_G [G,M]+
{\cal O}(\alpha_G^2),\\
&\psi\to e^{-i\alpha_G G} \psi=\psi-i\alpha_G \psi + {\cal O}(\alpha_G^2),
\end{split}
\ee
where $G$ denotes the representation of the baryon (B), third
component of the isospin (I) and hypercharge (Y) operators which are
related to the diagonal generators as $B=\sqrt{\frac{3}{2}}\lambda_0$,
$I=\frac{1}{2}\lambda_3$ and $Y=\frac{1}{\sqrt{3}}\lambda_8$.  The
coefficients in front of the diagonal matrices are chosen such as to
obtain the right quantum numbers when applying the operators on the
quark fields. The consequence of this symmetry is the existence of
conserved Noether vector-currents
\be
J_\mu^G=-\frac{\delta L}{\delta(\partial^\mu M)_{ij}}i[G,M]_{ji}
-\frac{\delta L}{\delta(\partial^\mu M^\dagger)_{ij}}i[G,M^\dagger]_{ji}
-\frac{\delta L}{\delta(\partial^\mu \psi_i)}iG_{ij}\psi_j.
\ee
The conserved charge is defined as $Q^G=\int d^3x J_0^G(x)$.
In terms of particle number operators the conserved baryon, isospin
and hypercharges read as
\bea
\label{Eq:Q_B}
Q^B&=&\frac{1}{3}(N_u+N_d+N_s-N_{\bar u}-N_{\bar d}-N_{\bar s}),\\
\label{Eq:Q_I}
Q^I&=&\frac{1}{2}(N_u-N_{\bar u}-N_d+N_{\bar d}+N_{\kappa^+}-N_{\kappa^-}
+N_{\bar\kappa^0}-N_{\kappa^0}+
N_{K^+}-N_{K^-}+
N_{\bar K^0}-N_{K^0})\nonumber\\
&&+N_{a_0^+}-N_{a_0^-}+N_{\pi^+}-N_{\pi^-},\\
\label{Eq:Q_Y}
Q^Y&=&\frac{1}{3}(N_u-N_{\bar u}+N_d-N_{\bar d}-2N_s+2N_{\bar s})+
N_{\kappa^+}-N_{\kappa^-}+
N_{\kappa^0}-N_{\bar\kappa^0}+
N_{K^+}-N_{K^-}+N_{K^0}-N_{\bar K^0}.
\eea
Note the different sign of $N_{K^0},N_{\bar K^0},N_{\kappa^0},
N_{\bar\kappa^0}$
in $Q^I$ relative to $Q^Y$. This is because  the particles 
$K^0,\bar K^0,\kappa^0,\bar\kappa^0$
fall into different
doublets from the point of view of the $I_3$ and $Y$ quantum numbers:
$K^0, K^+$ and $K^-, \bar K^0$ 
form a $I_3$ doublet while 
$K^0, K^-$ and $K^+,\bar K^0$
form a $Y$ doublet (likewise for scalars).

The statistical density matrix of the system is given by
\be
\rho=\exp[-\beta(H-\mu_G Q^G)],
\label{Eq:density_matrix}
\ee
with $G$ going over $B,I,Y$ in the summation over this index.
Using (\ref{Eq:Q_B}), (\ref{Eq:Q_I}), (\ref{Eq:Q_Y}) one can
rewrite (\ref{Eq:density_matrix}) by regrouping the terms in the
exponent according to different number operators and obtain
$\rho=\exp[-\beta(H-\mu_i N_i)],$ where $i$ goes over all the
non-singlet particles to which the following chemical potentials were
introduced in terms of $\mu_B,\mu_I,\mu_Y$:
\be
\begin{split}
&\mu_u=-\mu_{\bar u}=\frac{1}{3}\mu_{B}+\frac{1}{2}\mu_I+\frac{1}{3}\mu_Y,\\
&\mu_d=-\mu_{\bar d}=\frac{1}{3}\mu_{B}-\frac{1}{2}\mu_I+\frac{1}{3}\mu_Y,\\
&\mu_s=-\mu_{\bar s}=\frac{1}{3}\mu_{B}-\frac{2}{3}\mu_Y,\\
&\mu_{a_0^+}=\mu_{\pi^+}=-\mu_{a_0^-}=-\mu_{\pi^-}=\mu_{I},\\
&\mu_{\kappa^+}=\mu_{K^+}=-\mu_{\kappa^-}=-\mu_{K^-}=
\frac{1}{2}\mu_{I}+\mu_Y,\\
&\mu_{\kappa^0}=\mu_{K^0}=-\mu_{\bar\kappa^0}=-\mu_{\bar K^0}=
-\frac{1}{2}\mu_{I}+\mu_Y.
\end{split}
\label{Eq:all_the_chem_pots}
\ee

The singlet particles ($\pi^0,$ $\eta,$ $\eta',$ $a_0^0,$ 
$\sigma,$ and $f_0$) do not contribute to the conserved charges and in
consequence no chemical potential is introduced for them.  By looking
at (\ref{Eq:all_the_chem_pots}) one can see that different members
of a given multiplet ({\it e.g.} $\pi^+$ and $\pi^-$) acquire a
different combination of the baryon, isospin and hypercharge chemical
potentials, which means that the chemical potentials remove completely
the degeneracy between the members of the multiplets which we observe
in the vacuum, both at tree and one-loop level. We have to keep track
of the effect of 21 individually different particles, which makes
things more complicated than in previous studies of this model.

The effect of the chemical potentials is taken into account through
the propagators which are introduced using the definition familiar
from the theory of many-body systems. The relativistic formalism was
developed in \cite{zimmerman72} and is reviewed in
Appendix~\ref{ap:propagator}, where the calculation of the self-energy
using the finite-density Green's function is also sketched.

In order to see explicitly that the particle and its antiparticle
reflect differently the presence of a finite density medium we give
here the tree-level propagators of $K^+$ and $K^-$:
\be
\begin{split}
G_{K^+}(k)=\frac{i}{2 E_\k}\left[
\frac{1+n_{K^+}(E_\k)}{k_0-E_\k+i\epsilon}
-\frac{n_{K^+}(E_\k)}{k_0-E_\k-i\epsilon}
-\frac{1+n_{K^-}(E_\k)}{k_0+E_\k-i\epsilon}
+\frac{n_{K^-}(E_\k)}{k_0+E_\k+i\epsilon}
\right],\\
G_{K^-}(k)=
\frac{i}{2 E_\k}\left[
\frac{1+n_{K^-}(E_\k)}{k_0-E_\k+i\epsilon}
-\frac{n_{K^-}(E_\k)}{k_0-E_\k-i\epsilon}
-\frac{1+n_{K^+}(E_\k)}{k_0+E_\k-i\epsilon}
+\frac{n_{K^+}(E_\k)}{k_0+E_\k+i\epsilon}
\right],
\end{split}
\label{Eq:K+_K-_prop}
\ee
where $\displaystyle
n_{K^\pm}(E_\p)=\frac{1}{e^{\beta(E_\p-\mu_{K^\pm})}-1}$ and
$E_\p=\sqrt{\p^2+m_{K^{\pm}}^2}$.  The interpretation of the terms on
the right hand side of (\ref{Eq:K+_K-_prop}) is as follows (from left
to right): addition of a particle, removal of a particle, addition of
an antiparticle, removal of an antiparticle. Note, that in the
propagator of the $K^+$ the particle is $K^+$ and the antiparticle is
$K^-$, while in the propagator of the $K^-$ the particle is $K^-$ and
the antiparticle is $K^+$.

For all the other scalar and pseudoscalar fields the propagators can
be written analogously using the chemical potentials defined in
(\ref{Eq:all_the_chem_pots}). For the fermions the  propagators
are given in Appendix~\ref{ap:propagator}.

\section{Thermodynamics of the model at finite density
\label{sec:thermo}}

\subsection{The influence of $\mu_I$ and $\mu_Y$ on the CEP}

With the parameters fixed in the previous section, we can solve the
model at finite temperature and density using the formalism described
in Section \ref{chem_intro} and in Appendix
\ref{ap:propagator}. One calculates the 1-loop integrals entering the
finite temperature and density version of the equations which determine
the state of the system: the three equations of state
(\ref{Eq:eos_0}), (\ref{Eq:eos_3}), (\ref{Eq:eos_8}) and the
gap-equation for $m_{\pi^+}$ (\ref{Eq:pi+_gap_eq}). The relevant
integrals are given in Appendix \ref{ap:propagator}. An observed smooth
variation of the order parameters with the intensive parameter ($T$,
or $\mu_{B,I,Y}$) indicates analytic crossover type transition. A
first order phase transition is signaled by the multivaluedness of
either one of the three condensates in a given range of variation 
of the intensive parameter. The point where by varying some parameter(s) the
nature of the phase transition changes from crossover to first order
one corresponds to a second order phase transition.

\begin{figure}[t]
\includegraphics[keepaspectratio,width=1.0\textwidth,angle=0]{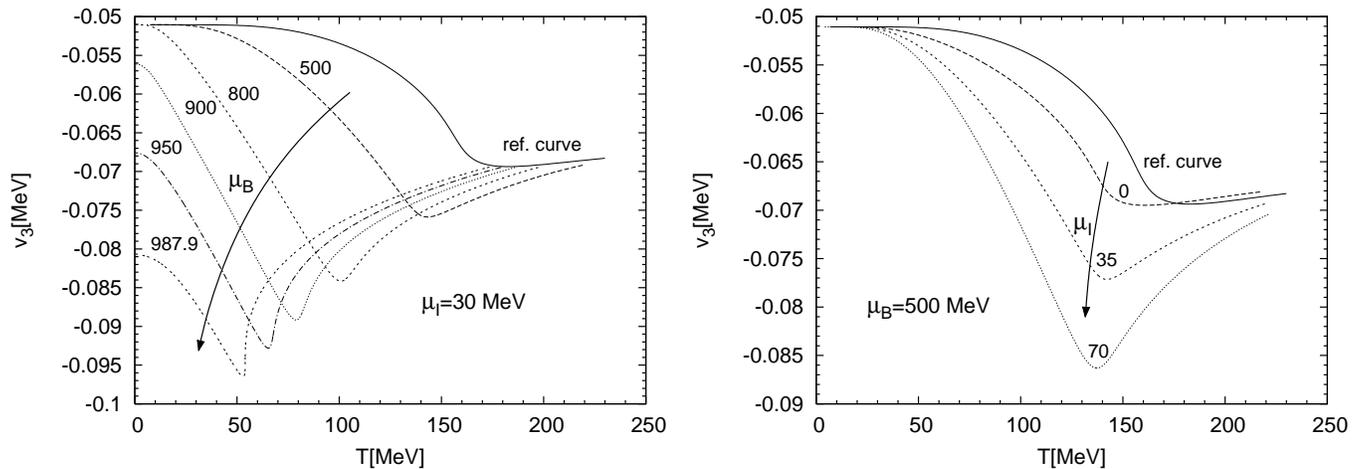}
\caption{The generic temperature dependence of $v_3$ for a crossover 
transition: on the l.h.s. $\mu_B$
changes and $\mu_I=70$~MeV, on the r.h.s.  $\mu_I$ changes while
$\mu_B=500$~MeV. On both panels the reference curve (ref. curve)
refers to the case $\mu_B=\mu_I=\mu_Y=0$.}
\label{Fig:vev}
\end{figure}

The critical end point (CEP) is a second order phase transition point
on the $\mu_B-T$ plane where by increasing $\mu_B$ the phase
transition as a function of $T$ changes from crossover to first order
($\mu_I$ and $\mu_Y$ are kept constant).
At vanishing $\mu_I$ and $\mu_Y$ the CEP is located in the point
$(T,\mu_B)_\textnormal{CEP}=(63.08,960.8)$~MeV. The
pseudo-critical temperature at vanishing chemical potentials is
$T_c(\mu_{B,I,Y}=0)=157.98$~MeV.

\begin{figure}[b]
\includegraphics[keepaspectratio,width=0.5\textwidth,angle=0]{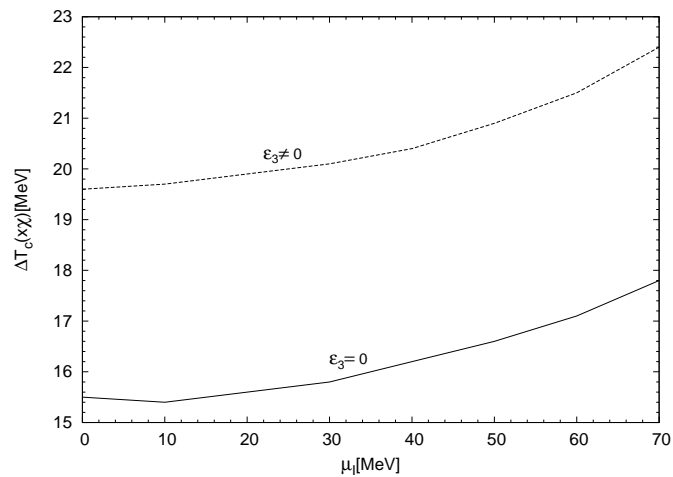}
\caption{ The width of the peak of the chiral susceptibility $\Delta
T_c(x\chi)$ as function of the isospin chemical potential with
(without) explicit symmetry breaking external field $\epsilon_3\ne 0$
($\epsilon_3=0$). In the chiral quark model $\chi=dx/d\epsilon_x$ where
$x=\sqrt{2/3}(v_0-v_8)$ is the non-strange condensate,
$\epsilon_x=\sqrt{2/3}(\epsilon_0-\epsilon_8)$ and as shown in
\cite{kovacs07} $\chi_{\bar\psi\psi}\sim x \chi$.  }
\label{Fig:width}
\end{figure}

Here it is important to note that, with the explicit isospin breaking
taken into account, these values have significantly changed with
respect to those obtained without isospin breaking at all (neither
explicit nor spontaneous):
$(T,\mu_B)_\textnormal{CEP}=(74.83,895.38)$~MeV and
$T_c(\mu_{B,I,Y}=0)=154.84$~MeV \cite{kovacs07}. At first sight this
is surprising since we have seen that at $T=\mu=0$ the effect of the
explicit symmetry breaking is minimal. The difference is due to the
behavior of the $v_3$ with the temperature. Without explicit isospin
symmetry breaking $v_3$ is identically zero for $\mu_I=0$. When
$\epsilon_3\ne0$ one can see by looking at the reference curve of
Fig.~\ref{Fig:vev} that with increasing temperature $v_3$ is
decreasing significantly compared to its $T=0$ value and reaches a
minimum around the phase transition point where the influence of $v_3$
becomes the strongest. The l.h.s. panel of Fig.~\ref{Fig:vev} shows
that the baryochemical potential magnifies this effect, implying that
approaching the CEP the influence of $v_3$ is even stronger. According
to our conjecture made in \cite{kovacs07} that a smoother crossover at
$\mu_B=0$ will require a larger value of $\mu_B$ to turn the
phase-transition in $T$ into a first order one, implying a larger
value of $\mu_{B,\textnormal{CEP}}$, we can expect that the larger
value of $\mu_{B,\textnormal{CEP}}$ in the case of the explicit
isospin breaking compared to the case in which the isospin breaking is
absent corresponds to a higher value of the width of the chiral
susceptibility $\Delta T_c(x\chi)$.  Indeed, by looking at
Fig.~\ref{Fig:width} one can see, that in the case with explicit
isospin symmetry breaking $\Delta T_c(x\chi)$ increased by $\sim
20\%$, approaching the value of $\Delta T_c(\chi_{\bar\psi\psi})=
28(5)(1)$~MeV at at $\mu_I=0$. This value was obtained on the lattice
in Ref.~\cite{fk06} after the extrapolation in the continuum limit was
done, though in this lattice investigation the effect of isospin
breaking was not taken into account. It would be interesting to see
whether similar effect is produced on lattice when $m_u\ne m_d$.

\begin{figure}[t]
\includegraphics[keepaspectratio,width=1.0\textwidth,angle=0]{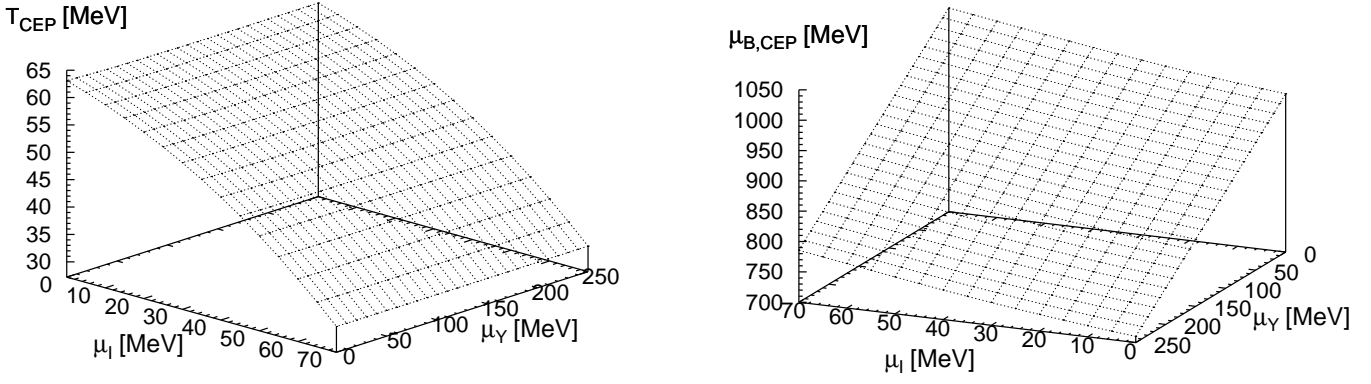}
\caption{The surfaces swept by the coordinates $T_\textnormal{CEP}$  and 
$\mu_{B,\textnormal{CE}P}$ of the critical end point as function of 
$\mu_I$ and $\mu_Y$.}
\label{Fig:CEP}
\end{figure}

Varying $\mu_I$ and $\mu_Y$ the location of the CEP in the $\mu_B-T$
plane changes. Fig.~\ref{Fig:CEP} shows the surfaces swept by the
two coordinates of the CEP
as functions of $\mu_I$ and $\mu_Y$. One can see that $\mu_Y$ has
practically no influence on $T_\textnormal{CEP}$, which decreases very
slowly, while with its increase $\mu_{B,\textnormal{CEP}}$ significantly
decreases. The increase of $\mu_I$ pushes the CEP towards higher
values of $\mu_{B,\textnormal{CEP}}$ and lower values of
$T_\textnormal{CEP}$. This behavior is in concordance to what was
previously written on the influence of $v_3$ on the CEP at $\mu_I=0$,
since by looking at the left hand side of Fig.~\ref{Fig:vev} one sees
that at finite $\mu_I$ the isospin condensate $v_3$ increases even
more with the temperature.

One can gain intuition on the way the chemical potentials $\mu_I$ and
$\mu_Y$ influence the coordinates of the CEP by attempting a simple
interpretation of our results in terms of generalized Clausius-Clapeyron
equations applied to our system. The particle number and entropy densities
of the two coexisting phases will be determined assuming an ideal gas of the
quasiparticle degrees of freedom, which differ only in their respective
masses on the two sides of the phase coexisting curves. The Clausius-Clapeyron
equation successfully describe the slopes of phase coexistence curves of
strong matter as functions of various chemical potentials and quark masses
\cite{halasz,nishida,gupta}. They are
derived from the Gibbs-Duhem relation which connect the variation of the
intensive thermodynamical parameters of a macroscopic system:
\be
\label{gibbs}
dp=s dT+n_B d\mu_B+n_I d\mu_I+n_Y d\mu_Y.
\ee
Here $n_B,n_Y,n_I$ are the particle number densities and 
$s$ is the entropy density. Keeping the pressure plus any other two
intensive parameters
constant one finds the following set of conditions for the phase
coexistence when one varies the remaining two intensive parameters along
the coexistence ``surface'':
\bea
\label{Eq:cc}
&{\displaystyle \frac {d T}{d \mu_B}\bigg|_{\mu_Y,\mu_I}=-\frac{\Delta n_B}{\Delta s},\quad
\frac {d T}{d \mu_Y}\bigg|_{\mu_B,\mu_I}=-\frac{\Delta n_Y}{\Delta s},\quad
\frac {d T}{d \mu_I}\bigg|_{\mu_B,\mu_Y}=-\frac{\Delta n_I}{\Delta
  s}}&,\nonumber \\
&{\displaystyle\frac {d \mu_B}{d \mu_Y}\bigg|_{T,\mu_I}=-\frac{\Delta n_Y}{\Delta n_B},\quad
\frac {d \mu_B}{d \mu_I}\bigg|_{T,\mu_Y}=-\frac{\Delta n_I}{\Delta n_B}.}&
\eea
On the right hand side of the equations above $\Delta$ refers to the 
difference of the values of a given extensive quantity in the symmetric and
broken symmetry phase. In the two coexisting phases the relevant particle 
number and/or entropy densities ($n_G$, $G=B,I,Y,$ and $s$) can be calculated 
from the partition function 
using the formulas $n_G=T V^{-1}\partial \ln Z/\partial \mu_G$ and 
$s=V^{-1}\partial (T\ln Z)/\partial T.$ Our simplified picture of the 
composition of the two phases in terms of non-interacting mixtures of 
15 quasiparticles is given by
\be
\ln Z=V\sum_i\gamma_i (2 s_i+1)\int\frac{d^3p}{(2\pi)^3}
\left[
\beta\omega_i+\ln(1+\alpha_i e^{-\beta(\omega_i-\mu_i)})
+\ln(1+\alpha_i e^{-\beta(\omega_i+\mu_i)})
\right], 
\ee 
where $i\in{\pi^{\pm},\pi^0,K^\pm,K^0,\eta,\eta',
a_0^\pm,a_0^0,\kappa^\pm,\kappa^0,\sigma,f_0,u,d,s}$, $\gamma_i=N_c$,
$\alpha_i=1$, $s_i=1/2$
for fermions and $\gamma_i=\alpha_i=-1$, $s_i=0$ for bosons, respectively. The
energies $\omega_i=\sqrt{p^2+m_i^2}$  are calculated with help of the
tree-level mass expressions (\ref{Eq:m_tree}) after substituting into them 
the order parameter values determined in our field theoretical treatment 
for the two phases, that is by solving (\ref{Eq:eos_0}), (\ref{Eq:eos_3}),
(\ref{Eq:eos_8}), and (\ref{Eq:pi+_gap_eq}).

The simple model predicts that 
$\Delta n_B, \Delta n_Y$ and $\Delta s$ is always positive, while
$\Delta n_I$ is always
negative. Moreover, the following relations are obtained: $\Delta n_B
\approx \Delta n_Y$, $\Delta s > \Delta n_B$ and
$\Delta s > |\Delta n_I|$. The discontinuity of the particle number
densities is determined by the contributions of essentially three
quasiparticles: $u, d,$ and $\pi^{\pm}.$ From our simple and transparent
model we get the sign and
even the magnitude of the shifts of the CEP in agreement with
Fig.~\ref{Fig:CEP} with the single exception of $d \mu_I/d T$.
The ideal
gas model does not reproduce the value of this derivative obtained by
solving our model. We interpret this discrepancy as a result of the
strong coupling between the 
$\langle \bar{u} u\rangle\sim \sqrt{2/3}v_0+\sqrt{1/3}v_8+v_3$ and
$\langle \bar{d} d\rangle\sim \sqrt{2/3}v_0+\sqrt{1/3}v_8-v_3$ 
condensates not captured by the ideal gas approximation. 
As one can check also in \cite{frank03}
the strong coupling between these condensates reduce 
the temperature of the CEP when a finite $\mu_I$ is switched on.
This is the same tendency we found in our field theoretical 
calculation. For the other three shifts it is the mass differences of
the lightest quasiparticles of the effective model which 
exert the strongest influence on the variation of CEP position.

\subsection{Quasi-particle masses}

We turn to the study of the dependence of the tree-level masses and
the one-loop pole masses on the temperature and the chemical
potentials. The one-loop pole masses are determined as the zeros of
the real part of the corresponding one-loop inverse propagators at
vanishing spatial momentum. For example, the equation determining the
one-loop $\pi^+$ mass reads:
$M_{\pi^+}^2=\textnormal{Re}G_{\pi^+}^{-1}(p_0=M_{\pi^+},\p=0)$.  If
there are more than one solutions of this type of equations, then we
follow that solution which in the vacuum lies closer to the physical
mass. Usually this solution is lost as the temperature increases and
some other solution is found.

\begin{figure}[h]
\includegraphics[keepaspectratio,width=0.95\textwidth,angle=0]{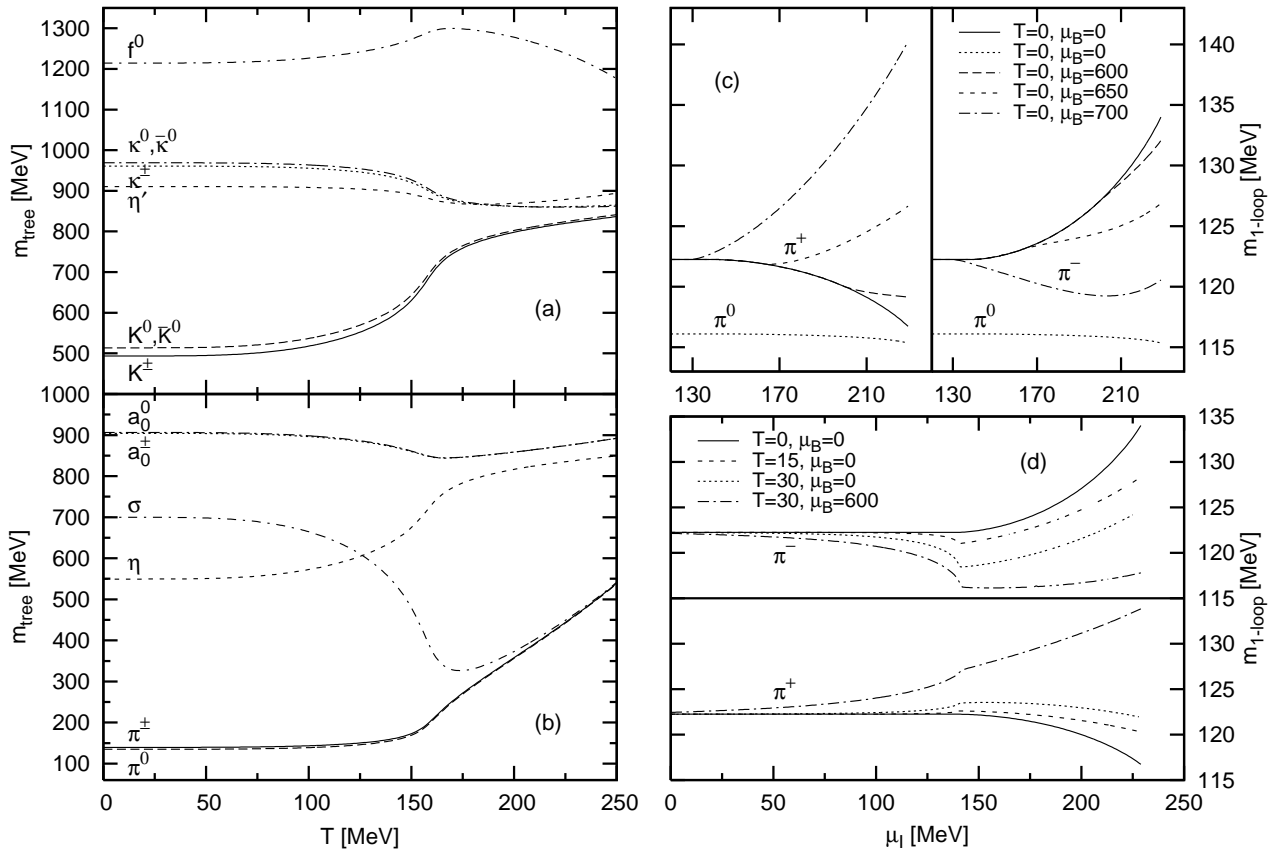}
\caption{The temperature dependence of the tree-level masses is shown
  in panels (a) and (b). The $\mu_I$ dependence of the one-loop pion
  masses for different values of $\mu_B$ at $T=0$ (panel (c)) and 
  $T\ne 0$ (panel (d)).}
\label{Fig:masses}
\end{figure}

In Fig.~\ref{Fig:masses}(a) and Fig.~\ref{Fig:masses}(b) we see that
the tree-level masses of $\pi^\pm,\pi^0$ and $\sigma$ clearly reflect
the restoration of the $SU(2)$ symmetry at high temperature. This is
not shown by the masses of $a_0^\pm, a_0^0$ and $\eta$. We cannot go
to higher values of the temperature because at $T\simeq 252$~MeV the
non-strange condensate $x$ decreases below the value of $v_3$ and the
tree-level mass of the $u$ quark turns into negative. At this
temperature there is still no sign for the tendency of the $SU(3)$
chiral partners to become degenerate.

In Fig.~\ref{Fig:masses}(c) we can see the dependence of the charged
and neutral pion masses on the isospin chemical potential. The charged pions
have by far the most significant dependence on $\mu_I$ from all of the
charged pseudoscalar mesons. At $T=\mu_B=0$ the splitting between
$\pi^+$ and $\pi^-$ is controlled by the bubble diagram involving
$\pi^+$ and $\pi^-$ respectively (see Fig.~\ref{Fig:self-energies})
and the splitting point is at $\mu_I\simeq m_\pi$. One can see that at
$T=\mu_B=0$ the mass of $\pi^0$ depends mildly on $\mu_I$. This
dependence intensifies with the increase of $T$ and $\mu_B$, but it
remains true that the dependence on $\mu_I$ is less strong than for
the case of $m_{\pi^\pm}$.  It is interesting to note that for large
values of $\mu_B$, when $\mu_{u/d}>m_{u/d}$ and the fermion bubble
contributes to the one-loop self-energies, the shape of the
$m_{\pi^\pm}(\mu_I)$ curves changes: $m_{\pi^+}$ starts to increase
with $\mu_I$ and the increase of $m_{\pi^-}$ with $\mu_I$ is slowed
down and eventually turned over into a decrease in a given interval of
$\mu_I$. Panel (d) shows that the increase of the temperature has a similar
effect as $\mu_B$ in that it turns over the $\mu_I$-dependence of
$m_{\pi^\pm}$ with respect to the behavior at $T=\mu_B=0$ starting at
a low value of $\mu_I$. 

\begin{figure}[ht]
\includegraphics[keepaspectratio,width=0.5\textwidth,angle=0]{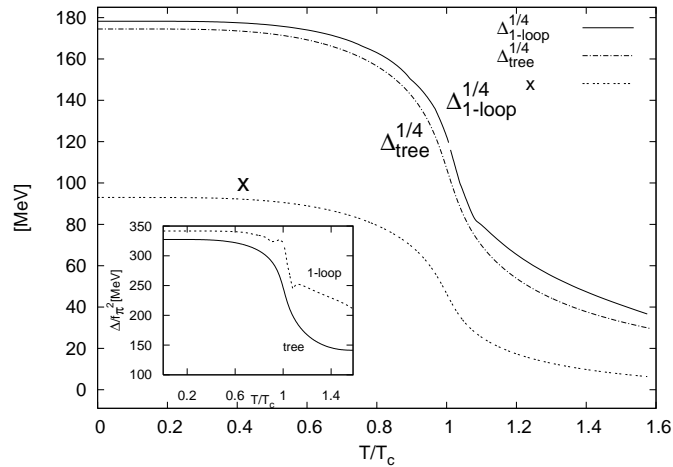}
\caption{
Estimation of the topological susceptibility  through the Witten--Veneziano
mass formula using one-loop and tree-level masses. The $T$-dependent pion
decay constant ($f_\pi(T)$) is approximated with the non-strange 
condensate $x$. The value of the pseudocritical temperature is
$T_c=157.98$~MeV. In the inserted figure $\Delta/f_{\pi}^2$ is plotted
based on the tree and one-loop masses.}
\label{Fig:WVMF}
\end{figure}

In Fig.~\ref{Fig:WVMF} we plot, both at tree and at one-loop level a
combination of the masses and the pion decay constant 
$\Delta=(m_\eta^2+m_{\eta'}^2-2m_K^2)f_\pi^2/6$, which through
the Witten--Veneziano mass formula \cite{Witten,Veneziano}
\be
\label{Eq:WV}
\frac{2N_f}{f_\pi^2}\chi_{{}_T}=m_\eta^2+m_{\eta'}^2-2m_K^2
\ee
with $N_f=3$, can be considered as an estimation of the topological
susceptibility $\chi_{{}_T}(T)$, which plays a crucial role in the
phenomenology of the $U(1)_A$ anomaly (see {\it e.g.}
\cite{costaUA1big,horvatic} for recent studies in terms of effective
descriptions).  In principle $\chi_{{}_T}(T)$ can also be computed
directly in our model if the quantity corresponding to the topological
charge density $Q_T$ of the QCD is extracted. This can be done by
comparing the four-divergence of the singlet axial vector current,
which in QCD involves the $U(1)_A$ anomaly term with the corresponding
current of the chiral quark model. Since the determinant term of
Eq.~(\ref{Lagrangian}) breaks the $U(1)_A$ symmetry, the
correspondence is $Q_T\sim g (\det(M)-\det(M^{\dag}))=g\textnormal{Im}
\det M$.

The decrease with $T$ of the estimated $\chi_{{}_T}(T)$ seen in
Fig.~\ref{Fig:WVMF} doesn't mean the restoration of the
$U(1)_A$ symmetry, since through $f_\pi(T)$, $\chi_{{}_T}(T)$ is
dominated by the restoration of the chiral symmetry. In view of
(\ref{Eq:WV}) this can be also seen on the inserted figure of
Fig.~\ref{Fig:WVMF}. However, the fact that at
$T=0$ the estimated $\chi_{{}_T}(T)$ is so close to the value obtained
on the lattice in \cite{Alles} and the curve itself stays within $10\%$ of the lattice
points, could imply that the effective restoration of the $U(1)_A$
symmetry, if contained in the lattice data
\footnote{Restoration of $U(1)_A$ symmetry requires that $\chi_{{}_T}(T)$
  decreases faster than $f_\pi(T)$ with the increase of $T$ so that 
$\chi_{{}_T}(T)/f_\pi^2(T)\to 0$.},
could be implemented in an effective description based
on the chiral quark model. In the NJL model the lattice result on
$\chi_{{}_T}(T)$ \cite{Alles} is converted into the temperature-dependence of
the strength of the determinant term, by fitting it with the explicit
formula of the susceptibility calculated in \cite{Fukushima}.

\section{Conclusions\label{sec:conclusion}}

In this paper we studied the influence of the isospin and hypercharge
chemical potentials on the $\mu_B-T$ chiral phase diagram of the three
flavored chiral constituent quark model with explicitly broken
$SU(3)_L\times SU(3)_R$ symmetry. The model was parametrized at one-loop
level and optimized perturbation theory was used for the resummation of the
perturbative series. Only one critical end point (CEP) is found for both
spontaneous and explicit isospin breaking.  In the latter case, based on the
width of the peak of the chiral susceptibility, the crossover transition at
$\mu_{B,I,Y}=0$ is found to be weaker than in the former case. Compared to
the case without isospin breaking, in the case with explicit isospin
breaking, the location of the CEP moves to a higher value of $\mu_B$ and a
lower value of $T$. For $\mu_I=\mu_Y=0$ the coordinates of the CEP are:
$(T,\mu_B)_\textnormal{CEP}=(63.08,960.8)$~MeV. This value of
$\mu_{B,\textnormal{CEP}}$ is about three times larger than the value found
on the lattice \cite{fk04} and increases (decreases) linearly with $\mu_I$
($\mu_Y$), while $T_\textnormal{CEP}$ is two fifth of the lattice value and
decreases slightly with the increase of $\mu_Y$ and significantly with the
increase of $\mu_I$. Using an ideal gas picture and the generalized
Clausius-Clapeyron equations we could interpret semiquantitatively with one 
exception the influence of $\mu_Y$ and $\mu_I$ chemical
potentials on the CEP as resulting from the quasiparticle masses. We also
studied the dependence of the charged and neutral one-loop pion masses on
the isospin chemical potential at different values of the temperature and
the baryon chemical potential. As a continuation of the present study, it
would be interesting to investigate at what value of the $\mu_I$ do the
charged pions condensate.

\section*{Acknowledgment}
Work supported by the Hungarian Scientific Research Fund (OTKA) under
contract number T046129, NI68228. Zs. Sz. is supported by OTKA Postdoctoral
Grant no. PD 050015. We thank A. Patk{\'o}s for discussion and
suggestions, especially on the correct way of introducing the various
chemical potentials, and for careful reading of the manuscript.

\appendix

\section{The formalism of relativistic many-body theory for 
a system at finite density and temperature \label{ap:propagator}}

We review below the method of relativistic many-body theory developed
in \cite{zimmerman72} for the perturbative calculation of the
self-energy at finite temperature and density.

First we present the derivation of the tree-level Green's functions for
$K^-,K^+,$ which depend both on the isospin and hypercharge chemical
potentials.  The field operators $K^-(x)$ and $K^+(x)$ are written in
terms of creation and annihilation operators $a^+(\p), b^+(\p)$ and
$a(\p), b(\p)$, respectively, as
\be
\begin{split}
&K^-(x)=\int\frac{d^3\p}{(2\pi)^3}\frac{1}{\sqrt{2E_\p}}\left(
a^+(\p)e^{ip\cdot x}+b(\p)e^{-ip\cdot x}
\right)\Big|_{p_0=E_\p},
\\
&K^+(x)=\int\frac{d^3\p}{(2\pi)^3}\frac{1}{\sqrt{2E_\p}}\left(
b^+(\p)e^{ip\cdot x}+a(\p)e^{-ip\cdot x}
\right)\Big|_{p_0=E_\p},
\end{split}
\label{Eq:field_K+_K-}
\ee
where $E_\p=\sqrt{\p^2+m_{K^{\pm}}^2}$.  This means that $a^+(\p)$
creates a $K^+$ particle, $b^+(\p)$ creates a $K^-$ particle, {\it etc}.
The operators have the usual non-zero commutators
\be
[a(\p),a^+(\k)]=[b(\p),b^+(\k)]=\delta(\p-\k).
\label{Eq:commutators}
\ee

The two point functions for $K^-$ and $K^+$ are defined as
\be
\begin{split}
&G_{K^-}(y-x):=\lag T K^-(y)K^+(x) \rag_\beta
=\Theta(y_0-x_0)\lag K^-(y)K^+(x)\rag_\beta+
\Theta(x_0-y_0)\lag K^+(x)K^-(y)\rag_\beta,\\
&G_{K^+}(y-x):=\lag T K^+(y)K^-(x) \rag_\beta
=\Theta(y_0-x_0)\lag K^+(y)K^-(x)\rag_\beta+
\Theta(x_0-y_0)\lag K^-(x)K^+(y)\rag_\beta,
\end{split}
\label{Eq:K+_K-_prop1}
\ee
where the average is to be taken over a grand canonical ensemble, that
is for an operator $O$ one has
\be
\lag O\rag_\beta=\frac{\tr [e^{-\beta{\cal H}} O]}
{\tr e^{-\beta{\cal H}}},
\ee
with ${\cal H}=H-\mu_i Q_i$. We make this distinction between $K^+$
and $K^-$ propagators because the particle and its antiparticle feel
differently the presence of the dense medium, resulting in a different
mass dependence on the chemical potential. In our case this
difference in the mass manifests itself first at one-loop level.

Substituting (\ref{Eq:field_K+_K-}) into (\ref{Eq:K+_K-_prop1}),
taking only the non-interacting part of the Hamiltonian $H$, with the
help of the commutator relations given in (\ref{Eq:commutators}) and
the Campbell--Baker--Hausdorff relation one evaluates the expectation
values obtaining
\be
\lag a^+(\p)a(\q)\rag_\beta=\delta(\p-\q) n_{K^+}(E_\p),\qquad
\lag b^+(\p) b(\q)\rag_\beta=\delta(\p-\q)\bar n_{K^+}(E_\p),
\label{Eq:a+a}
\ee
where $\displaystyle n_{K^+}(E_\p)=\frac{1}{e^{\beta(E_\p-\mu_{K^+})}-1}$,
$\displaystyle \bar n_{K^+}(E_\p)=\frac{1}{e^{\beta(E_\p+\mu_{K^+})}-1}$.
Note, that $\bar n_{K^+}(E_\p)=n_{K^-}(E_\p)$.
Using (\ref{Eq:a+a}) and the Fourier representation of $\Theta(t)$ in
(\ref{Eq:K+_K-_prop1}) one obtains in momentum space the $K^+$ and
$K^-$ propagators given in (\ref{Eq:K+_K-_prop}).

Next, we calculate a one-loop bosonic bubble appearing in
Fig.~\ref{Fig:self-energies}. 
With the standard rules of the perturbation theory, using the
conventions of \cite{peskin_book} the $\pi^+$ self-energy is given by
\be
-i\Sigma_{\pi^+}(y,x)=-4(3\tilde G_{2\beta\gamma}+
4 \tilde H_{2\beta,\gamma \delta} v_\delta)
(3\tilde G_{1\beta'\gamma'}+
4 \tilde H_{1\beta',\gamma' \delta'} v_{\delta'}) 
G_{\pi_{\beta'}\pi_{\beta}}(y,x) 
G_{\sigma_{\gamma'}\sigma_{\gamma}}(y,x).
\ee

The first non-mixing bubble graph in the diagrammatic representation of 
$\Sigma_{\pi^+}$ given in Fig.~\ref{Fig:self-energies} is obtained
with the choice $\beta=4, \beta'=5$ implying $\gamma=7, \gamma'=6$. Using
that $\tilde G_{156}=\tilde G_{247}$ and 
$\tilde H_{15,6\delta}=\tilde H_{24,7\delta}$ the contribution of this
graph is
\be
-i\Sigma_{\pi^+}^{K^+ \bar \kappa^0}(y,x)=\!\parbox{3.3cm}{
\includegraphics[keepaspectratio,width=0.18\textwidth,angle=0]{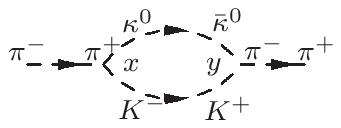}}
\!=-4\left[
3\tilde G_{247}+4\tilde H_{24,7\delta}v_\delta\right]^2
G_{K^+ K^-}(y,x) G_{\bar{\kappa}^0\kappa^0}(y,x).
\ee 
The labels in the graph denote the field operators, {\it e.g.}
on the left hand side $\pi^-$ creates a $\pi^+$ particle. 

Going to momentum space one has
\be
\Sigma_{\pi^+}^{K^+ \bar \kappa^0}(p)=-4iV_{\pi^+ K^-\kappa^0}^2
\int \frac{d^4k}{(2\pi)^4} G_{K^+}(k) G_{\bar\kappa^0}(p-k)=
4V_{\pi^+ K^-\kappa^0}^2
I^{\beta}_B(p,m_{K^+},\mu_{K^+},m_{\bar\kappa^0},\mu_{\bar\kappa^0}),
\ee
where the vertex is $V_{\pi^+ K^-\kappa^0}=4\left[
\frac{c}{\sqrt{2}}+\frac{g_2}{\sqrt{3}}v_0
-\sqrt{2}g_2\left(\frac{v_8}{\sqrt{3}}-v_3\right)\right]$, and
$G_{K^+}(k)\equiv G_{K^+ K^-}(k)$.

Generally, at finite chemical potentials and temperature for a bosonic
bubble diagram one calculates at vanishing spatial external momentum
($\p=0$) an integral of the form:  
\bea
\nonumber
I^{\beta}_B(p_0,m_1,\mu_1,m_2,\mu_2)&=&
-i\int\frac{d^4 k}{(2\pi)^4}G_1(k) G_2(p-k)\bigg|_{\p=0}\\
&=&\hspace*{-0.2cm}\int\frac{d^3 \k}{(2\pi)^3}\frac{1}{4E_1E_2}\left[
\frac{1+n_1+n_2}{p_0-E_1-E_2}-\frac{n_1-\bar n_2}{p_0-E_1+E_2}
+\frac{\bar n_1-n_2}{p_0+E_1-E_2}-\frac{1+\bar n_1+\bar n_2}{p_0+E_1+E_2}
\right],
\label{Eq:bubble_integral}
\eea
where for the propagators one uses a form similar to that in 
(\ref{Eq:K+_K-_prop})
and to arrive at the second equality one performs a contour
integration in the complex energy plane.
The distribution functions $n_i\equiv n_i(E_i)$ with
$E_i=\sqrt{\k^2+m_i^2}$ 
contain the chemical potential for particle or 
antiparticle which is created by the fields 
of the vertex in the left hand side.

We rewrite the integral (\ref{Eq:bubble_integral}) as:
\bea
I_B^{\beta}(p_0,m_1,\mu_1,m_2,\mu_2)&=&I_B^{\mu,T=0}(p_0,m_1,m_2)\no\\
&&+\frac{1}{8\pi^2p_0}\sum_{i=1}^2\Pval\int_{m_i}^{\infty}dE\sqrt{E^2-m_i^2}
\left[\frac{n_i(E)}{p_0a_i-E}+\frac{\bar{n}_i(E)}{p_0a_i+E}\right],
\eea
where the remaining integral is evaluated numerically, $\Pval$ stands
for principal value.
The vacuum integral $I_B^{\mu,T=0}(p_0,m_1,m_2)$ is given by the
expression (B4) of \cite{kovacs07},
$n_i=1/(\exp(\beta(E-\mu_i))-1)$ is the Bose-Einstein 
distribution and 
$a_i=[1+(-1)^{i-1}(m_1^2-m_2^2)/p_0^2]/2$.

For fermions the method is identical to that used for the bosons.
The fermion propagators for the constituent quarks $u,\bar u$ are defined as
\be
\begin{split}
&D_u(y-x):=\lag T u(y)\bar u(x) \rag_\beta
=\Theta(y_0-x_0)\lag u(y)\bar u(x)\rag_\beta-
\Theta(x_0-y_0)\lag \bar u(x)u(y)\rag_\beta,\\
&D_{\bar u}(y-x):=\lag T \bar{u}(y) u(x) \rag_\beta
=\Theta(y_0-x_0)\lag \bar{u}(y) u(x)\rag_\beta-
\Theta(x_0-y_0)\lag u(x)\bar{u}(y)\rag_\beta,
\end{split}
\ee
which in the momentum space read
\be
\begin{split}
&D_u(k)=
\frac{i(\slk+m_u)}{2 E_\k}\left[
\frac{1-f_u^+(E_\k)}{k_0-E_\k+i\epsilon}
+\frac{f_u^+(E_\k)}{k_0-E_\k-i\epsilon}
-\frac{1-f_u^-(E_\k)}{k_0+E_\k-i\epsilon}
-\frac{f_u^-(E_\k)}{k_0+E_\k+i\epsilon}
\right],\\
&D_{\bar u}(k)=
\frac{i(\slk+m_u)}{2 E_\k}\left[
\frac{1-f_u^-(E_\k)}{k_0-E_\k+i\epsilon}
+\frac{f_u^-(E_\k)}{k_0-E_\k-i\epsilon}
-\frac{1-f_u^+(E_\k)}{k_0+E_\k-i\epsilon}
-\frac{f_u^+(E_\k)}{k_0+E_\k+i\epsilon}
\right],
\end{split}
\ee 
where $\displaystyle f^+_u(E_\p)=\frac{1}{e^{\beta(E_\p-\mu_u)}+1}$
and $\displaystyle f^-_u(E_\p)=\frac{1}{e^{\beta(E_\p+\mu_u)}+1}$
are the distribution functions for $u$ type quarks and antiquarks.

Then for the fermionic bubble appearing in the $\pi^+$
self-energy (see  Fig.~\ref{Fig:self-energies}) one has
\be
\Sigma_{\pi^+}^{u \bar d}(p)=-\frac{g_F^2}{2}N_c i \tr
\int \frac{d^4k}{(2\pi)^4} \gamma_5 D_{\bar d}(k) D_{u}(k+p)=
\frac{g_F^2}{2}N_c
I^{\beta}_F(p,m_d,\mu_{\bar d},m_{u},\mu_{u}).
\ee
Similarly to equation (\ref{Eq:bubble_integral}) in case of fermions
we use the integral:
\bea
I^{\beta}_F(p_0,m_1,\mu_1,\mu_2,m_2)&=&-i\tr\int_k \gamma_5
D_1(k)\gamma_5 D_2(k+p)\bigg|_{\p=0} \no \\
&=&\int\frac{d^3
  \k}{(2\pi)^3}\left[\frac{1}{E_1}(f_1^++f_1^--1)+\frac{1}{E_2}(f_2^++f_2^--1)\right] \no\\
+2(p_0^2-(m_1-m_2)^2)&&\hspace*{-0.75cm}\int\frac{d^3 \k}{(2\pi)^3}\frac{1}{4E_1E_2}\left[
\frac{1-f_1^+-f_2^+}{p_0-E_1-E_2}+\frac{f_1^+-f_2^-}{p_0-E_1+E_2}
-\frac{f_1^--f_2^+}{p_0+E_1-E_2}-\frac{1-f_1^--f_2^-}{p_0+E_1+E_2}
\right]\no\\
&=& -2T_F^{\mu,T=0}(m_1)-2T_F^{\mu,T=0}(m_2)+
2(p_0^2-(m_1-m_2)^2)I_B^{\mu,T=0}(p_0,m_1,m_2)\no\\
+T_F^{T\neq0}(m_1)\hspace*{-0.18cm}&+&\hspace*{-0.18cm}T_F^{T\neq0}(m_2)-\frac{p_0^2-(m_1-m_2)^2}{4\pi^2p_0}
\sum_{i=1}^2\Pval\int_{m_i}^{\infty}dE\sqrt{E^2-m_i^2}\left[\frac{f^+_i(E)}{p_0a_i-E}+
\frac{f^-_i(E)}{p_0a_i+E}\right]
\eea
where $f^{\pm}_i=1/(\exp(\beta(E\mp\mu_i))+1)$ is the Fermi-Dirac 
distribution.

}
\end{document}